\documentstyle[12pt,epsf,epsfig,wrapfig]{article} 
\begin{document}
\setcounter{footnote}{0}
\begin{flushright}
{\tt hep-ph/9710411}
\end{flushright}
\bigskip
\begin{center}
{\large \bf
NUCLEAR STRUCTURE FUNCTIONS AND CUMULATIVE PROCESSES}\footnote
{This is an updated version of the talk~\cite{efr0} at 13 International
Seminar on Relativistic Nuclear Physics and QCD (Dubna, Sept. 2-7, 1996)
which includes new experimental data and some
ideas appeared since that time. Partialy supported by the RFBR Grant 
96-02-17631.}\\
\vspace{5mm}
{\bf A.V. Efremov}   \\
\vspace{2mm}
{\small\it
JINR, Dubna, 141980 Russia  }
\end{center}

\vspace{2mm}
\begin{center}
ABSTRACT

\vspace{5mm}
\begin{minipage}{130 mm}
\small
The authors's point of view, based on QCD, on the nuclear quark structure is
presented.  Different models for explaining the EMC--effect are considered.
It is also shown that cumulative production data are very useful for a better
understanding of the EMC--effect and give some evidence in favor of its
multiquark nature.
\end{minipage}
\end{center}
\setcounter{equation}{0}
\setcounter{figure}{0}
\setcounter{table}{0}

Discovery of the EMC--effect~\cite{efr1} has drawn attention of the
world--wide community of physicists to the problem of quark structure of
nuclei, and to its irreducibility to the quark structure of constituent
nucleons only~\cite{efr2}. Stream of theoretical papers followed the
discovery of EMC suggesting a whole spectrum of possibilities for
understanding the phenomena~\cite{efr3}. However, many of the suggestions
met with difficulties after a change of experimental data on $F_A/F_D$ in
the region of small $x$~\cite{efr4,efr5}. Nowadays, when all suggestions
seem to be made, one can try to analyze them on a general basis and to
estimate to what extent the nuclear quark structure is understood and what
is still unclear.

\medskip
{\large\bf 1. Connection of nucleon and nuclear quark structure.}
Probably G.West first noticed~\cite{efr6} that QCD evolution equations result in a
simple convolution relation of {\it nonsinglet} quark distribution
functions (the
valence quarks) of nucleus and nucleon
\footnote{Since the evolution
equations do not
depend on the kind of object
$$
\dot V_A(n,Q^2)/V_A(n,Q^2)= \dot V_N(n,Q^2)/V_N(n,Q^2)=\gamma_N(\alpha_S(Q^2))
$$
(the dot means derivative with respect to $\log Q^2$ and $n$ is the number of
a moment) the first equality immediately gives $V_A(n,Q^2)=
T_A^{NS}(n)V_N(n,Q^2)$
which is equivalent to (\ref{efr1a}).}

\begin{equation}
xF_{3A}\approx V_A(x,Q^2)=
\int_x^Ad\alpha\, T_A^{NS}(\alpha)V_N\left({x\over\alpha},Q^2\right)
\label{efr1a}
\end{equation}
where the function $T_A^{NS}(\alpha)$ satisfies the baryon number sum rule
\begin{equation}
\int_0^Ad\alpha\, T_A^{NS}(\alpha)=1
\label{efr1b}
\end{equation}
(all nuclear function here are normalized to A). Due to this, one can
consider $T_A^{NS}(\alpha)$ as an effective ''valence nucleon'' distribution
function over a fraction of momenta $\alpha$ in spite of the
impossibility of expressing it through a one--nucleon wave function.

A similar relation can be written for the singlet channel as
well~\cite{efr7,efr8} which mixes the singlet quark, $\Sigma(x,q^2)=
\sum_q x[q(x,Q^2)+\bar q(x,Q^2)]\approx F_2 $,
and gluon, $G(x,Q^2)$, distribution functions
\begin{eqnarray}
\Sigma_A(x,Q^2)=\int_x^Ad\alpha\, T_A^{S}(\alpha)
\Sigma_N\left({x\over\alpha},Q^2\right)
\label{efr2a}           \\
G_A(x,Q^2)=\int_x^Ad\alpha\, T_A^{S}(\alpha)G_N\left({x\over\alpha},Q^2\right)
\label{efr2b}
\end{eqnarray}
where, in general, $T^S_A\neq T_A^{NS}$ and $T_A^S$ satisfies the
energy--momentum sum rule
\begin{equation}
\int_0^Ad\alpha\,\alpha T_A^{S}(\alpha)={M_A\over AM_N}\approx 1
\label{efr2c}
\end{equation}

Really, diagonalizing the system of two linear evolution equations
for the moments $\Sigma_A(n,Q^2)$ and $G_A(n,Q^2)$, one can obtain the relation
for two eigenfunctions $f^{\pm}(n,Q^2)=\Sigma(n,Q^2)+
C_n^{\pm}(\alpha_S(Q^2))G(n,Q^2)$
($C_n^{\pm}$ are some diagonalizing coefficients depending on
anomalous dimension matrix elements):
\begin{equation}
f_A^{\pm}(n,Q^2)= T^{\pm}_A(n)\,f_N^{\pm}(n,Q^2)\ .
\label{efr3}
\end{equation}

The quark (and gluon) distribution function is expressed through the limit
of quark (or gluon) propagator $\langle P|\bar q(0)q(\xi)|P\rangle$ when
$\xi\to 0$, regularized with the help of an ultraviolet cutoff parameter
$Q^2$. It must satisfy Bethe-Salpeter equations (Fig. 1) with
inhomogeneous terms. The second equation for the gluon propagator however
become
{\it homogeneous}, at least in the so--called leading logarithm
approximation, i.e.
disregarding $\alpha_S^n(\log Q^2)^{n-1}$ corrections. In this
approximation also, it became an {\it algebraic} equation for the moments of
structure functions with coefficients which are {\it independent} of a
target. This means that the ratio $\Sigma(n)/G(n)$ is also
independent on the target and as a result of (\ref{efr3})
$T_A^+=T_A^-=T_A^S$ which leads to the relations (\ref{efr2a},\ref{efr2b}).

Physically this approximation corresponds to a rather widely accepted
scheme in which the gluon distribution is totally due to QCD
evolution process starting from the valence quark distribution at
low $Q^2$, i.e. when there is no "primordial" gluon distribution
(see e.g. the work~\cite{efr8a}).

\begin{figure}[bh]
\centering
\small
\mbox{\epsfig{figure=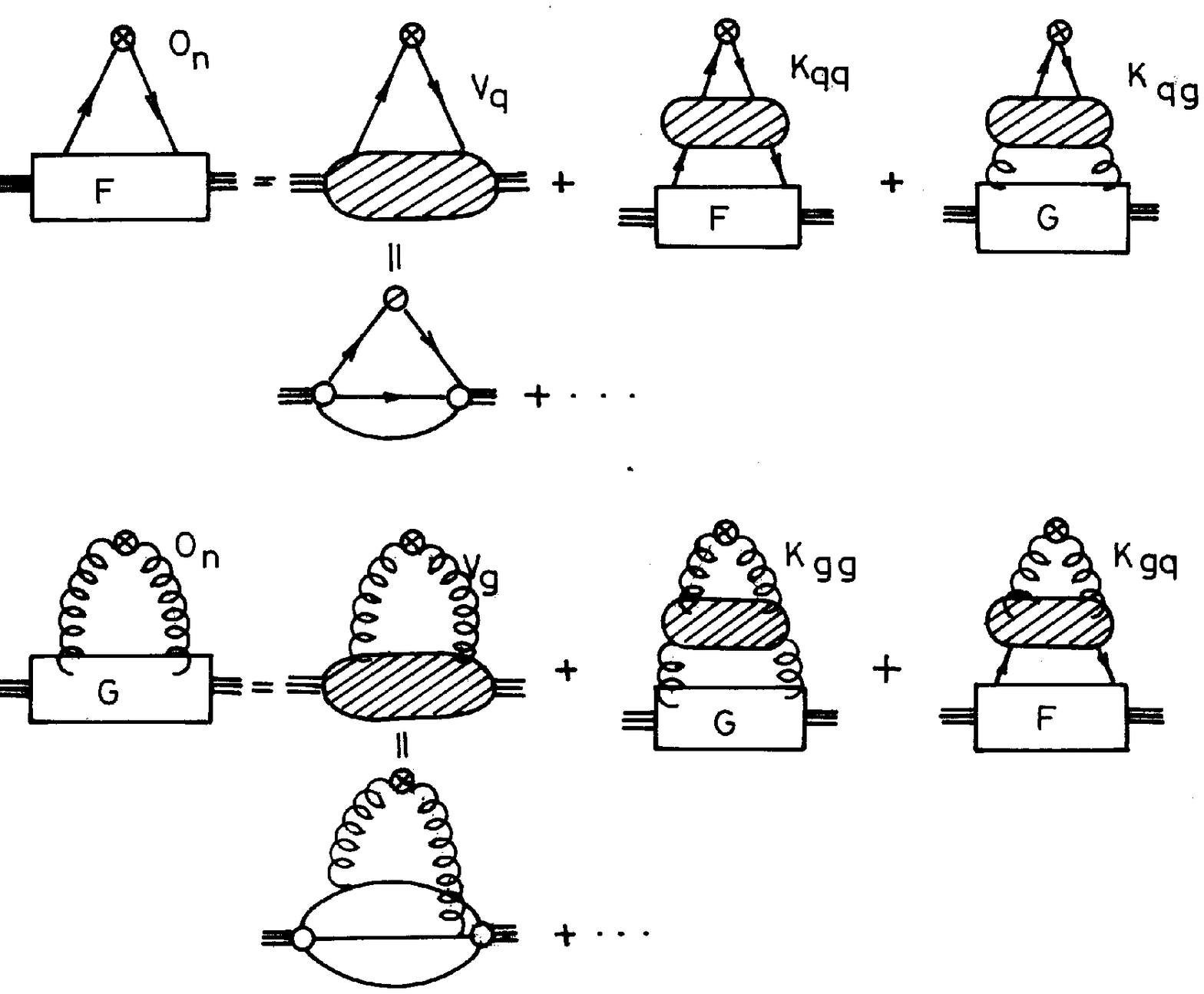,width=12cm,height=6cm}} \\
{\footnotesize Fig. 1. Bethe-Salpeter equation for singlet quark and gluon
structure functions.}
\label{fig1}
\end{figure}

An immediate consequence of Eqs. (\ref{efr2a},\ref{efr2b}) is the
{\it equality} of average momenta fractions of gluons and quarks in the
nucleus and nucleon
\begin{equation}
{<x_G>_A\over<x_G>_N}= {<x_q>_A\over<x_q>_N}
\label{efr4}
\end{equation}
This relation is in good agreement with BCDMS~\cite{efr4} data which are
the most precise nowadays, e.g.  $<x_q>_{N_2}/<x_q>_{D_2}-1 =
(0.7\pm1.7\pm1.0)\%$. The difference of average momenta for other nuclei
is also zero within the error bars (see Table 2 in Ref.~\cite{efr8a}).

The relation (\ref{efr4}) clearly contradicts the very popular rescaling
hypothesis~\cite{efr9} in explanation of the EMC--effect. In fact, the
passage from nucleon to nucleus in these models is equivalent to the growth
of $Q^2$ for which, according to QCD, $<x_G>$ increases and $<x_q>$
{\it decreases}\footnote{Another criticism of the hypothesis from QCD point
of view can be found in Ref. \cite{efr7}.}.

In conclusion of this section let us stress once more that QCD evolution
equations just as relation (\ref{efr3}) are results of the leading twist
approximation. So, the relations (\ref{efr1a}) and
(\ref{efr2a},\ref{efr2b}) do not include the nuclear screening which is, at
least formally, a high--twist effect~\cite{efr10,efr11}. Some experimental
observation of a significant $Q^2$--dependence of $F_{Sn}/F_C$ in the
region $x<0.02$ was known recently~\cite{efr11a}.

\medskip
{\large\bf 2. The EMC--effect.}
Let us see now what the EMC--effect means in the frame of our approach.
Let us assume that the functions $T_A$ determine, at least approximately,
an effective distribution of nucleons in nucleus and therefore they are
mostly concentrated in the region of $\alpha=1$ (i.e. in the region of zero
internal momentum of a nucleon). Expanding $F_N({x\over\alpha})$ in
(\ref{efr1a}) and (\ref{efr2a}) around $\alpha=1$, it is easy to obtain for
not very large $x$
\begin{equation}
R={F_A\over F_N}\simeq <T>+<(1-\alpha)T>\,x{F'_N\over F_N}+
{1\over2}<(1-\alpha)^2T>\,x\left(x{F^{''}_N\over F_N}+2{F'_N\over F_N}\right)
+\cdots
\label{efr5}
\end{equation}
where $<\cdots>$ means integration over interval $0<\alpha<A$. If one
accept that $F_N\sim (1-x)^k$ and $k\simeq 3$, then $x$-dependence of the
second and the third terms are the factors $-k[x/(1-x)]$ and
$k[x/(1-x)]\cdot[(k-1)x/(1-x)-2]$ respectively. In the region of
$x\approx0.5$ the second term is close to zero and to obtain the depletion
of $R$ from unity in this region discovered by EMC one should have
\begin{equation}
<T^S_A>-1=\Delta_A>0 \qquad\mbox{ or }\qquad
\int_0^Ad\alpha\,\left[T^S_A(\alpha)-T^{NS}_A(\alpha)\right]=\Delta_A>0
\label{efr6}
\end{equation}
for the ratio $R_2$ of the structure functions $F_2\approx\Sigma$ and
\begin{equation}
1-<\alpha T^{NS}_A>=\delta_A>0 \qquad\mbox{ or }\qquad
\int_0^Ad\alpha\,\alpha\left[T^S_A(\alpha)-T^{NS}_A(\alpha)\right]=\delta_A>0
\label{efr7}
\end{equation}
for the ratio $R_3$ of the structure functions  of $xF_3$.

In addition, in the region $x\approx0.5$ the sea quarks are practically
absent: therefore one can expect here $R_3\simeq R_2$ and
\begin{equation}
\delta_A\simeq\Delta_A \qquad\mbox{ (more exactly ${2\over3}\Delta_A$)}
\label{efr8}
\end{equation}

The relations (\ref{efr6}) and (\ref{efr7}), mean that the number of
"effective nucleons" in a nucleus has to be {\it larger} than $A$, and
the valence nucleons have to carry only {\it a part} of the total nucleus
momentum. In other words, the EMC--effect is the result of a {\it repumping}
of a part of momentum from valence quarks to sea quarks in the nucleus
in comparison with free nucleon.

Notice, that the shock produced by the discovery of EMC was due to the
prejudice that a nucleus is made of $A$ nucleons and so the condition
$\Delta_A=0$ has to be imposed on the distribution $T^S$, which unavoidably
results in $R_2(x\approx 0.5)=1$, {\it independent} of the form of $T^S$.
In this sense, the difference between $T^S$ and $T^{NS}$ (necessary to
explain the EMC--effect) leads to the {\it irreducibility} of the nuclear
quark structure to the quark structure of free nucleons.

In spite of generality, this approach allows one to draw a number of
interesting conclusions:

\smallskip
{\it i}) It immediately follows from (\ref{efr6}) that the ratio
\footnote{Recall that the screening phenomena are disregarded here}
\begin{equation}
R_2(x\simeq 0)=\int_0^Ad\alpha\,T^S_A(\alpha)=1+\Delta_A>1
\label{efr9}
\end{equation}
The most accurate measurement of BCDMS~\cite{efr4} shows a small $\approx
5\%$ but definite excess of the ratio over 1 in the region of small $x$,
i.e. the same value as the loss of momenta of the valence nucleons
$\delta_A$.  This means a small number of particles of the non-nucleon
component which have to be {\it heavy} enough to supply the 5\% repumping
of the momentum (e.g. $\rho$-mesons, $N\widetilde N$--pairs or pions far
off the mass shell).

\smallskip
{\it ii}) In addition to the internucleon sea there
is a small, $\approx\Delta_A$, but {\it hard} enough ''collective sea''
of quark--antiquark pairs in nuclei.

Using (\ref{efr1a}) and (\ref{efr2a}) it is easy to obtain for the sea
\begin{eqnarray}
&&O_A(x,Q^2)\equiv\Sigma_A-V_A
\nonumber                                         \\
&=&\int_x^Ad\alpha\,T_A^{NS}(\alpha)
O_N\left({x\over\alpha},Q^2\right)+\int_x^Ad\alpha\,\left[T_A^{S}(\alpha)-
T_A^{NS}(\alpha)\right]\Sigma_N\left({x\over\alpha},Q^2\right)
\label{efr10}
\end{eqnarray}
where the first term comes from the internucleon sea, which
rapidly decreases with increasing $x$, and the second term comes
from a "collective sea", $O'_A$, which is hard since its center of gravity is
\begin{equation}
\overline\alpha_{O'}={<\alpha(T^S_A-T^{NS}_A)>\over
<T^S_A-T^{NS}_A>}={\delta_A\over\Delta_A}\approx 1
\label{efr11}
\end{equation}
For pions on the mass shell this number is $m_{\pi}/m_N\approx 1/7$.
That is the reason why
the repumping of the momentum into the pions~\cite{efr12} gives no satisfactory
description of the data in the region of small $x$ (too many pions are
needed to supply the 5\% repumping).

\smallskip
{\it iii}) The place of intersection $R_2(x_0)=1$ {\it does not depend} on the
sort of nucleus and is at $x_0\approx0.3$. Really, if there are no
screening and light particles in nuclei, $T_A^S(\alpha)$ has to be smooth
enough in the region of small $x$. Using then the first two terms of
(\ref{efr5}) for $R_2$ it is easy to find
$$
{x_0\over 1-x_0}\simeq\frac{1}{3}\left(1-
{\int_0^{x_0}d\alpha\,\alpha T^S_A(\alpha)\over
\int_0^{x_0}d\alpha\,T^S_A(\alpha)}\right)^{-1}
$$
The ratio of integrals in the r.h.s. is in the interval
$[0,x_0]$ and thus $0.25<x_0<1/3$. This feature of the ratio $R_2$ has been
well confirmed experimentally~\cite{efr12a} with $x_0=0.278\pm0.010$.

Now, what about different models proposed? They are, in fact,
different suggestions of the repumping mechanisms. Not all of them
seem satisfactory from our viewpoint. We have mentioned the rescaling
models~\cite{efr9} where part of the repumping comes into gluon
component. However, the main drawback of these models is the {\it softness}
of the gluon and the sea component in nucleon. This leads to an extra big
value of $R_2(0)$ after the 5\% repumping. (Although the authors deny the
applicability of their model to the region of small $x$.) As it was
noticed, models with repumping of momenta into the mass shell
pions~\cite{efr12} have the same disadvantage.

Other models can be divided into three big categories:

\smallskip
{\it i}) Models with repumping of the momentum either into massive meson
component~\cite{efr13} ($\rho,\ \omega$, off the mass--shell pions) or into
nucleon--antinucleon pairs~\cite{efr8}. A component like that is probably
related to the core of nuclear force at small distances. However, it is
hard to believe that the nucleon can conserve at such small distances its
individual quark structure without converting it into multiquark states.

\smallskip
{\it ii}) Repumping inside each nucleon~\cite{efr14}, i.e. change of its quark
structure due to the influence of the internuclear field.  Transition of
part of nucleons into $\Delta$-isobar ~\cite{efr15} also belongs to this
class. We do not see, however, how it is possible to obtain the hard sea
here.

\smallskip
{\it iii}) Repumping inside a multiquark fluctuation~\cite{efr16}. By this
we mean not only a bound state but also a state of two or more nucleons
with interaction of their quarks, as proposed in ~\cite{efr10}, or with an
exchange quark interaction in the final state considered in~\cite{efr17}.
That kind of interactions is inevitable in any theory with a composite
nucleon. However, the calculation of the quark structure of the states like
that seems as difficult as the calculation of the quark structure of
nucleus. Recently some progress in this direction has been
achieved~\cite{efr16a}.

\medskip
It is necessary to stress the important difference between a multiquark
state and few-nucleon correlation (FNC)~\cite{efr18}. The loss of momenta
of the valence quarks for the latter are the same as averaged over the
nucleus, $\Delta_{FNC}=\Delta_A$, due to a change of structure of {\it
each} nucleon. For the multiquark, however, it has to be much larger
\begin{equation}
\Delta_{6q}>\Delta_A
\label{efr12}
\end{equation}
e.g. if there is no repumping inside the nucleons, then
$\Delta_A=p_A\Delta_{6q}$, where $p_A$ is a probability of multiquark
states. In fact, the relation (\ref{efr12}) can be considered as a {\it
definition} of the multiquark state in distinction with FNC.  A statistical
realization of the hard antiquark sea is also known (see e.g. Kondratyuk
paper~\cite{efr16}).

It seems that structure function measurements alone cannot distinguish
between these models. So, new sources of information are necessary.
One of them is deep inelastic scattering with measurement of hadrons
in a final state. Production of $\rho$- and $\Delta$-resonances and also
$K^-$-mesons and antiprotons which carry the information about the collective
sea is especially interesting for evident reasons.

\medskip
{\large\bf 3. Cumulative particles production.} Another source of
information is the cumulative particle production.  Especially, the
production of $K^-$-mesons and antiprotons off nuclei in the region $x\geq
1$, because of the peculiarity of the nuclear quark structure mentioned
before.


A question arises however: to what extent is the cumulative production cross
section determined by the nuclear structure functions $F_A(x)$?  Until
now there are no quite reliable data for nuclear deep inelastic
scattering in the region $x\geq 1$, though there are some indications of
similarity of the cumulative meson spectra and structure function $F_2(x)$
in this region~\cite{efr20}.

There exist two points of view on the physics of cumulative
production~\cite{efr2}: (a) ''hot models'', in which massive clusters in
nuclei (which are necessary to produce a cumulative particle) are {\it formed}
by an incoming hadron either by a sort of compression of the nuclear
matter and heavy fireball formation or by multiple rescattering~\cite{efr20a};
(b) ''cold models'', in which formations of that sort {\it already exist}
in nuclei because of Blokhintsev's fluctuations of density~\cite{efr21}
either in a form of multiquark states or in a form of a few nucleon
correlation, resulting in the high momentum tail of Fermi motion. This
reflects in the structure functions of the nucleus. A common property of
these models is the independence of the nuclear parton fragmentation of
the nucleus type. This allows us to write down the cross section of the
process in the form
\begin{equation}
{\epsilon\over A}\,{d\sigma\over d^3p}=\rho_{A\to h}(x,y,p_T)=
\int_x^A{d\alpha\over\alpha}\,F_A(\alpha)f_h\left({x\over\alpha},y,p_T\right)
\label{efr13}
\end{equation}
where $x=-u/s$, $y=-t/s$ and the function $f_h$ does not depend on $A$, i.e.
it is the same for a nucleus and for a free nucleon. Combining (\ref{efr13})
with (\ref{efr1a},\ref{efr2a}), it is easy to obtain a natural expression
\begin{equation}
\rho_{A\to h}(x,y,p_T)=
\int_x^Ad\alpha\,N_A(\alpha)\rho_{N\to h}\left({x\over\alpha},y,p_T\right)+
\int_x^Ad\alpha\,\widetilde N_A(\alpha)
\rho_{\widetilde N\to h}\left({x\over\alpha},y,p_T\right)
\label{efr14a}
\end{equation}
where we use the notation
\begin{equation}
N_A={1\over2}\left(T^S_A+T^{NS}_A\right) \quad\mbox{ and }\quad
\widetilde N_A={1\over2}\left(T^S_A-T^{NS}_A\right)\ .
\label{efr14b}
\end{equation}

The first expression in (\ref{efr14b}) can be considered approximately, due
to smallness of the EMC--effect, as a distribution of nucleons over
fractions of the momentum. For cumulative and stripping protons it is
necessary to add to (\ref{efr14a}) a term proportional to $N_A(x)$ which
takes into account dissociation of the nucleus. Moreover, just this term
gives a major contribution when $p_T\approx0$~\cite{efr23}. Parametrizing
the form of the spectrum of stripped and cumulative protons with
$p_T\approx0$ (with normalization $<N_A>=1+\Delta_A/2$, $<\alpha
N_A>=1-\delta_A/2$) and using the experimental cross section for
$\rho_{N\to\pi}$, one obtains the cross section of cumulative pion
production without any new parameter. (The second term in (\ref{efr14a})
naturally gives a small correction.) This program for deuterium (to
minimize possible secondary nuclear effects) has been made in
work~\cite{efr22} and shows  a good agreement with experiment. Also, the
ratio $K^+/\pi^+$ agrees with experiment. This agreement confirms the
independence of fragmentation of the kind of a nucleus (at least, for light
nuclei), which is the base of (\ref{efr14a},\ref{efr14b}) and means also
that the valence mesons carry the same information on the nuclear quark
structure as the cumulative protons~\cite{efr24}.  However, the main
peculiarity of the nuclear quark structure is hidden here.

Interpretation of $\widetilde N_A$ in (\ref{efr14a},\ref{efr14b}) depends
on the mechanism of repumping and, due to the  second term in
(\ref{efr14a}), dominates for ''sea particles'' $K^-,\widetilde p$ in the
region $x\geq 1$. They are just sensitive to the peculiarity of the nuclear
quark structure. For the ratio of $K^+/K^-$ yields in the region we have
\begin{equation}
r_A={K^+\over K^-}\approx
{\int_x^Ad\alpha\,N_A(\alpha)\rho_{N\to K^+}({x\over\alpha})\over
\int_x^Ad\alpha\,\widetilde N_A(\alpha)
\rho_{N\to K^+}({x\over\alpha})}\ ,
\label{efr15}
\end{equation}
where the approximation $\rho_{\widetilde N\to K^-}\approx\rho_{N\to K^+}$
is used.  It is known experimentally~\cite{efr25,efr25c,efr25a} that the
ratio $r_A$ for various nuclei ($Be,\ Al,\ Cu,\ Ta,\ Pb$) is {\it constant}
in $x$, within experimental accuracy, in the region $1<x<2.5$ (Figs. 2)
\footnote{ Notice that the variable $x$ here is defined by a minimal mass
(in the nucleon mass units) of $M_X^2=(xp_A+p_B-p_C)^2$ for the process
$B+A\to C+X$, where $p_A$ is a 4-momentum of a nuclei per a nucleon.}
is in {\it sharp contradiction} with its behaviour in region $x<1$.
Therefore, the  functions $N_A$ and $\widetilde N_A$ in this region should
only differ by a factor. One can assume that this difference reflects
the difference in normalization condition for this functions
\begin{equation}
<\alpha\widetilde N_A>\approx<\widetilde N_A>=\Delta_A/2 \quad\mbox{ and }
<\alpha N_A>\approx<N_A>\approx 1
\label{efr16}
\end{equation}

So for the models of type i) and ii) one would expect from (\ref{efr15})
$r_A\approx2/\Delta_A$. Using the parametrization~\cite{efr8} of the
SLAC~\cite{efr26} and BCDMS~\cite{efr4} data for the EMC--effect one finds
$r_{Be}=86$ ($\Delta_{Be}=0,023$), $r_{Al}=55$ ($\Delta_{Al}=0.036$),
$r_{Ta}=36$ ($\Delta_{Ta}=0.056$) and $r_{Pb}=35$ ($\Delta_{Pb}=0.058$),
which is {\it significantly higher} than the experimental ratio (see
Figs. 2), especially for light nuclei.

\begin{figure}[bh]
\small
\raisebox{1ex}{\mbox{\epsfig{figure=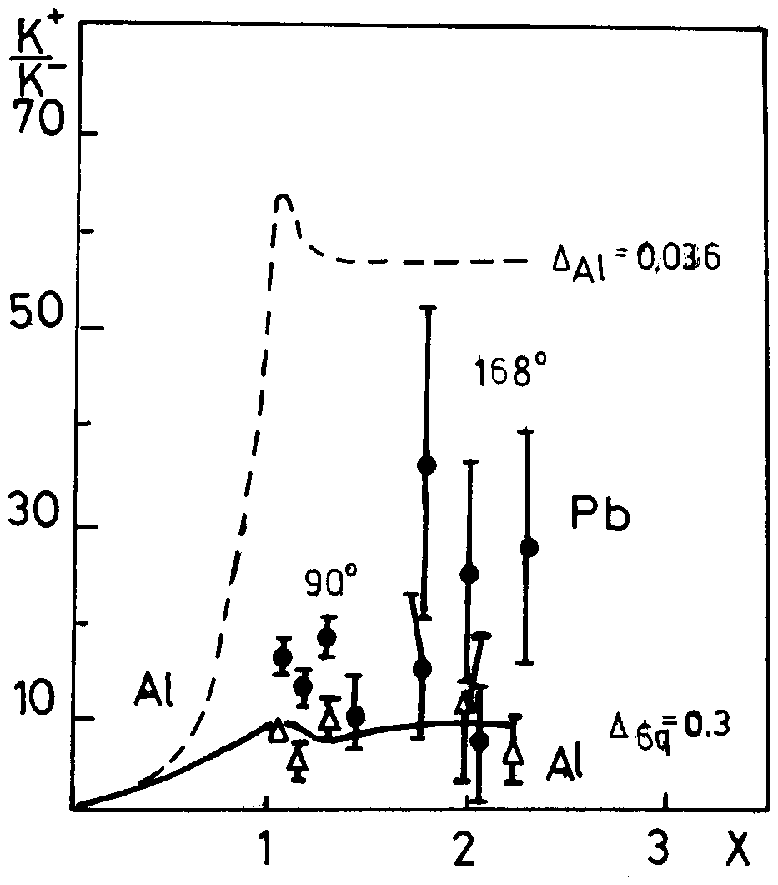,width=4.7cm,height=4.7cm}}}
\raisebox{-1.3ex}{\mbox{\epsfig{figure=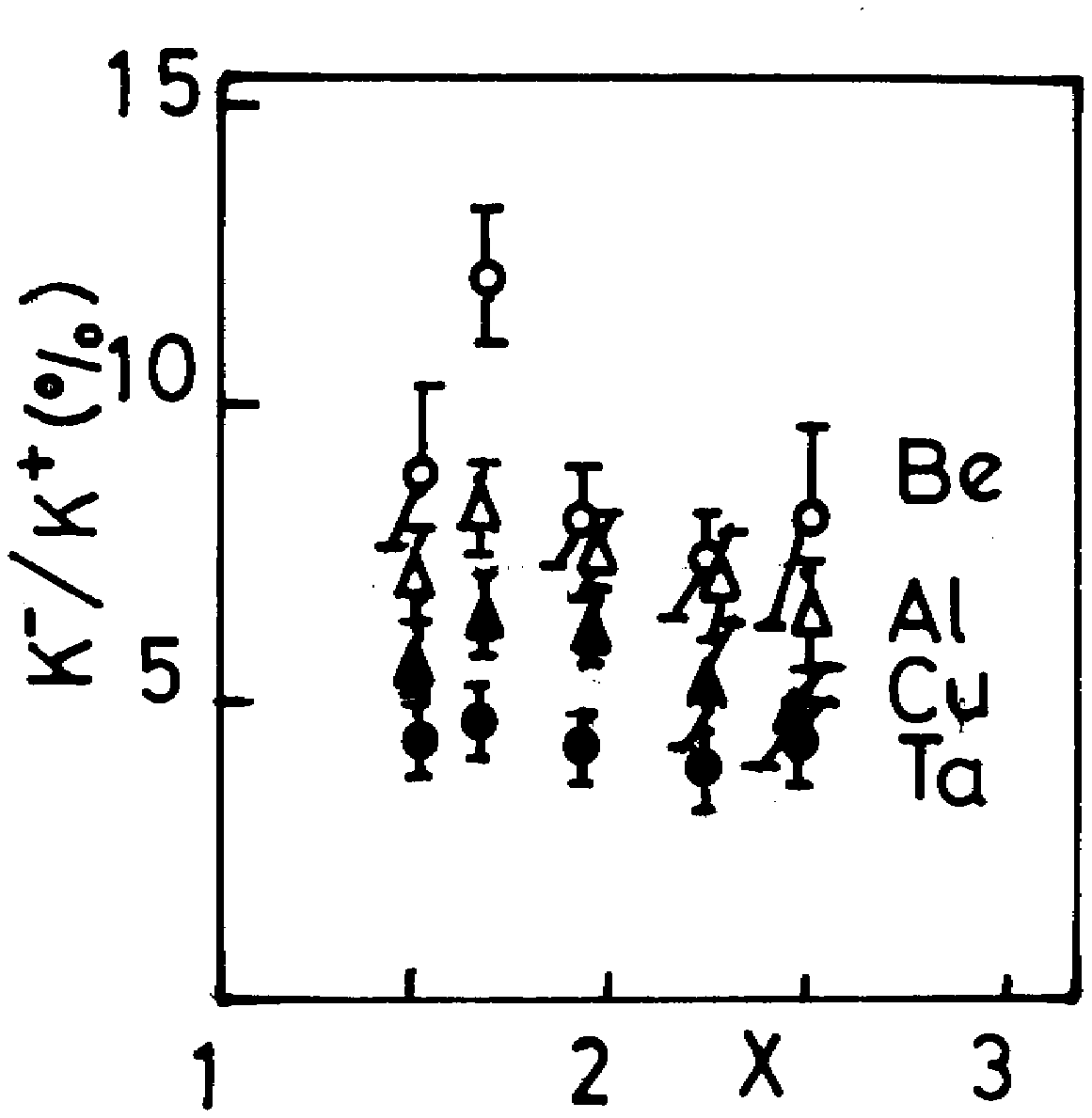,width=5.3cm,height=5.3cm}}}
\raisebox{-0.8ex}{\mbox{\epsfig{figure=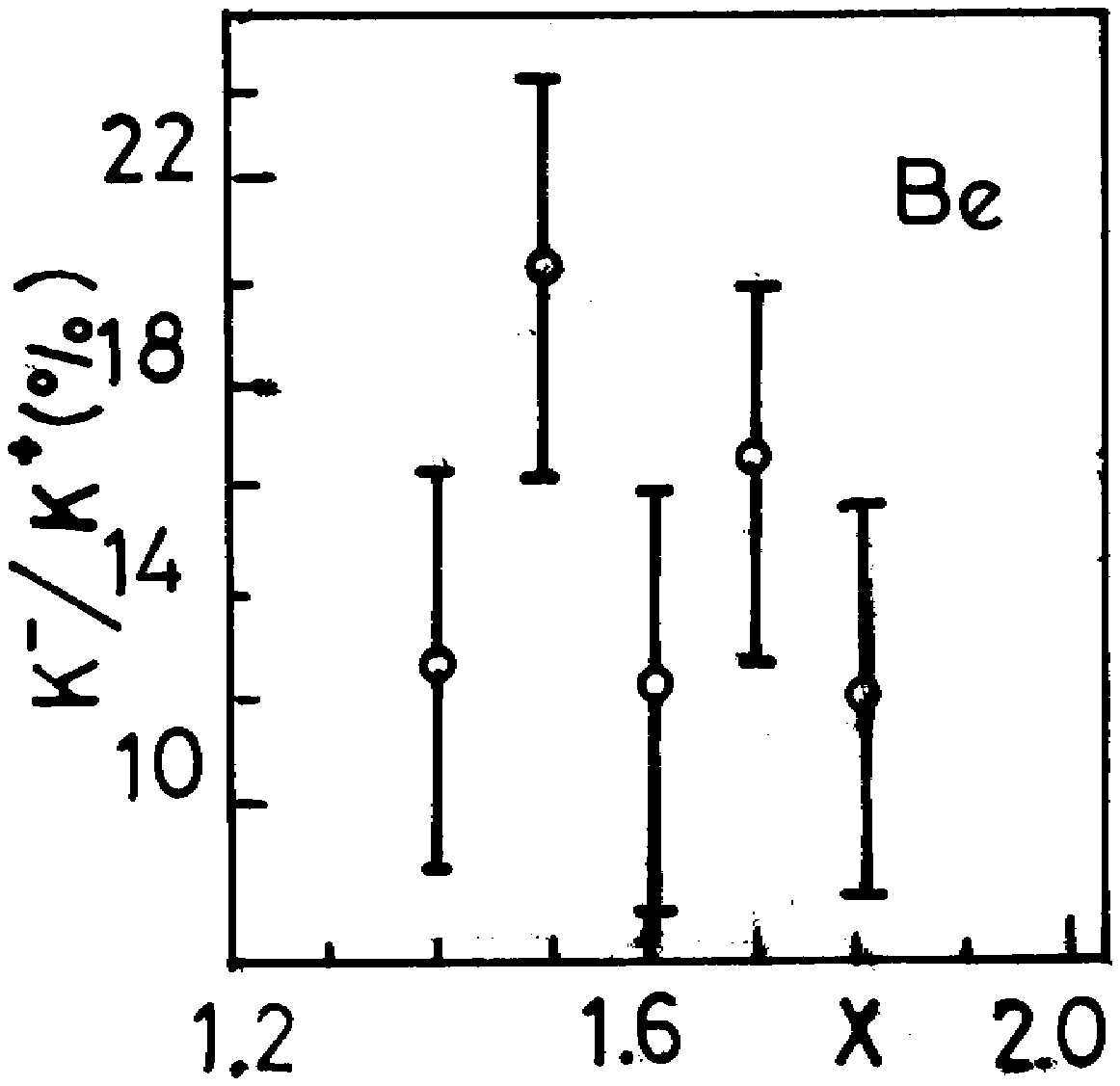,width=5.5cm,height=5.6cm}}} \\
{\footnotesize Fig.2. The relative yields of cumulative $K^+$ and $K^-$
from works~\cite{efr25} ($E_{lab}=8.9\,GeV,\; \theta_{lab}=168^\circ$),
\cite{efr25c} ($10\,GeV,\; 119^\circ$) and \cite{efr25a} ($40\,GeV,\;
159^\circ$).}
\label{fig234}
\end{figure}

\begin{wrapfigure}{R}{5.5cm}
\mbox{\epsfig{figure=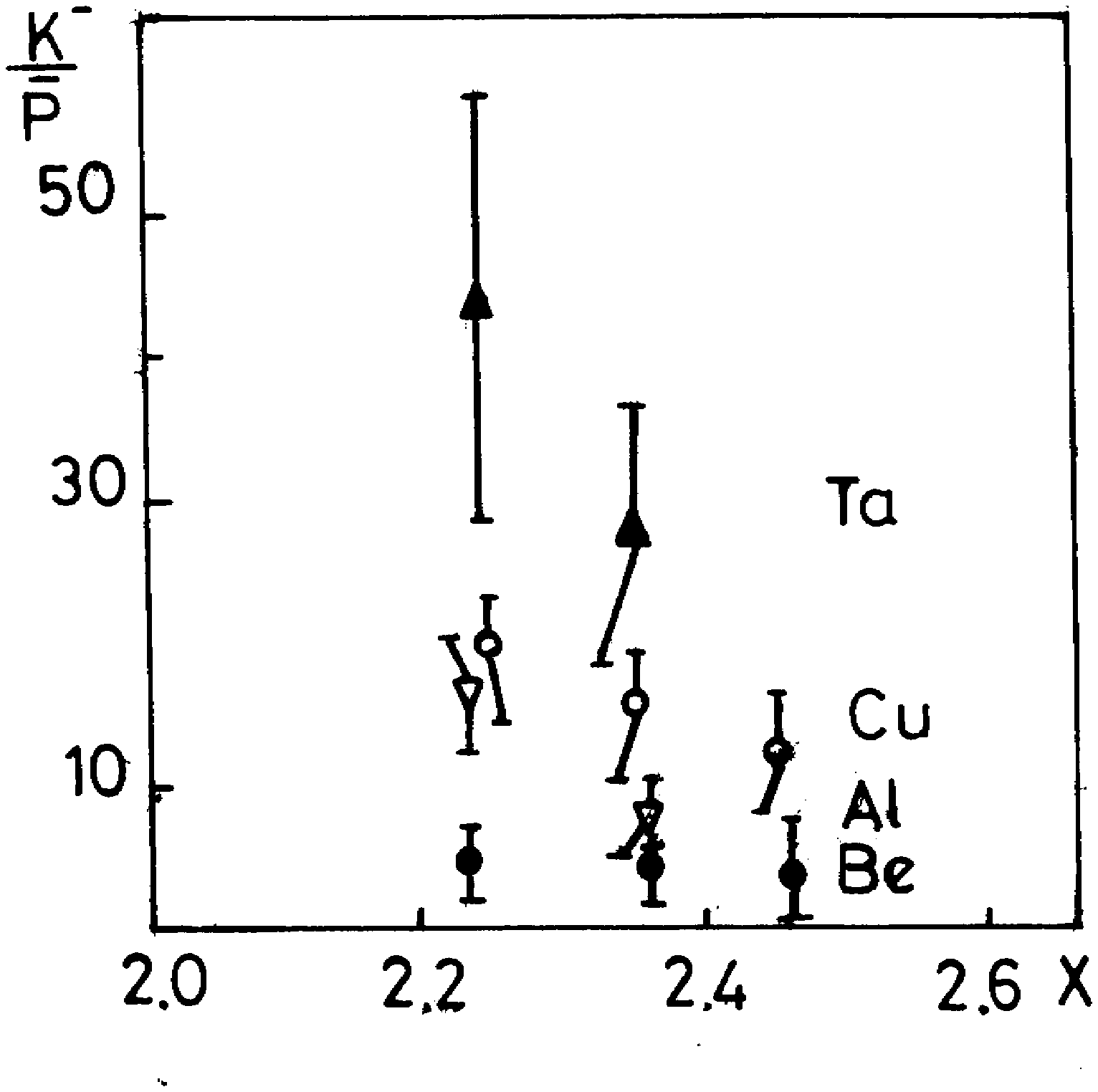,width=5.5cm,height=5.5cm}} \\

\vspace{-8mm}
{\footnotesize Fig.3. The $K^-/\widetilde p$ yields ratio from
works~\cite{efr27} ($10\,GeV,\; 119^\circ$).}
\label{fig5}
\end{wrapfigure}

For the repumping inside multiquark states, which has to determine the
cumulative cross sections in this region of $x$, the repumping
$\Delta_{6q}\simeq\Delta_A/p_A$ has to be larger (due to a small $p_A$) and
$r_A\simeq2/\Delta_{6q}$ has to be lower. E.g. the experimental ratio
$r^{exp}_{Be}\simeq 9$ corresponds to $\Delta_{6q}\simeq0.22$ and
$p_{Be}\simeq0.10$.  So, the low ratio of $r_A$ can be considered as an
indication of the multiquark mechanism of both cumulative phenomena and the
EMC--effect.

Now, let us turn to the cumulative antiprotons. Naturally, they are
sensitive to the $N\widetilde N$-pair repumping mechanism~\cite{efr8}.  The
ratio of $p/\widetilde p$ yields is determined by an expression of the type
of (\ref{efr15}) and is of the order $2/\Delta_A\simeq 10^2$.  The
experiment~\cite{efr27} gives for this ratio $p/\widetilde p\approx10^8$,
which definitely rejects the above mechanism.  On the other hand, with no
packing of the collective sea into $N\widetilde N$-pairs, the cumulative
$\widetilde p$ can arise in fragmentation of $\widetilde q$ (just as
$K^-$). Then the ratio $\widetilde p/K^-$ has to be $\simeq0.1$
(suppression by an order of magnitude is due to fragmentation $\widetilde
q\to\widetilde p$ and some growth due to a smaller transverse momentum of
$\widetilde p$ at the same value of $x$) which is not far from the
experimental observation~\cite{efr27} $K^-/\widetilde p\simeq 5$ for $Be$
(see Fig. 3\ref{fig5}).

More accurate calculations of $K^+/K^-$ and $\widetilde p/K^-$ ratios
based on the dual string model with the multi-quark fluctons were done in
paper~\cite{efr28}. Notice, however, that the model underestimates the
$K^+/K^-$--ratio for the proton target in the region $x>0.6$.

It is necessary to stress also that secondary nuclear effects can be
significant for the intermediate and heavy nuclei we have considered.
Indications to these effects come, for example, from an enhanced
$A$-dependence of cumulative proton and $K^+$ productions and from an
enhanced by 4--5 times depletion from unity of
$\rho_{A\to\pi}/\rho_{D\to\pi}$ in the region $x\approx0.6$ compared to the
deep inelastic scattering.  (Notice also that the ratio of cumulative cross
sections $He/D$ shows even an {\it anti}--EMC effect in this region!) For
these reasons it would be very desirable to have accurate data on the
kaon and antiproton production off deuterium.

\medskip
The conclusive headlines are:

\smallskip
{\it i}) The cause of the EMC--effect is the repumping of the valence quark
momentum to a collective sea of quark--antiquark pairs.

\smallskip
{\it ii}) Small excess of the $A/D$ ratio in the $x\approx 0$ region points to
a hardness of a collective sea or to a massive non-nucleon component
in nuclei.

\smallskip
{\it iii}) Many popular models are in trouble due to {\it i}) and {\it ii}).

\smallskip
{\it iv}) The ratio of $K^+/K^-$ cumulative cross sections supports the
multiquark mechanism of the EMC--effect and of the cumulative process.

\smallskip
{\it v}) Accurate data on the kaon and antiproton production off deuterium are
highly desirable.


\begin{thebibliography}{99}
\bibitem{efr0} Efremov A.V. Tashkent Symp. on Multiparticle Dynamics.
Singapore PC, 1987, p. 689.
\vspace{-3mm}
\bibitem{efr1} Aubert J.J. et al. (EMC) Phys.Lett. {\bf B123} (1983) 275.
\vspace{-3mm}
\bibitem{efr2} Baldin A.M. PANIC X--th Int. Conf., Heidelberg,
1984, p.J11; Progr. Part.Nucl.Phys. {\bf4} (1980) 95,
Pergamon Press; Stavinski V.S. Elem.Part. and Nucl. (EPAN) {\bf 11}
(1980) 571; Efremov A.V. EPAN {\bf13} (1982) 613;
Blokhintsev D.I. Proc. of 19-th Int. Conf. on HEP, Tokyo, 1978, p.475.
\vspace{-3mm}
\bibitem{efr3} Krzywicki A. Nucl.Phys. {\bf A446} (1985) 135.
\vspace{-3mm}
\bibitem{efr4} Benvenuti A.C. et al. (BCDMS) Phys.Lett.
{\bf B189} (1987) 483.
\vspace{-3mm}
\bibitem{efr5} Norton P.R. Proc. of 23-th Int. Conf. on HEP,
Berkley, 1986.
\vspace{-3mm}
\bibitem{efr6} West G.B. Los Alamos Prepr. LA--UR 84-2072 (1984).
\vspace{-3mm}
\bibitem{efr7} Efremov A.V. Yad.Fiz. {\bf 44} (1986) 776.
\vspace{-3mm}
\bibitem{efr8} Efremov A.V. Phys.Lett. {\bf B174} (1986) 219.
\vspace{-3mm}
\bibitem{efr8a} Barone V., Genovese M., Nikolaev N.N., Predazzi E.,
Zakharov B.G. Z.Phys. {\bf C58} (1993) 541.
\vspace{-3mm}
\bibitem{efr9} Close F.E., Roberts R.G., Rose G.G. Phys. Lett.
{\bf B129} (1983) 346;
Close F.E., Jaffe R.L., Roberts R.G., Ross G.G., Phys.Rev.
{\bf D31} (1985) 1004.
Nachtmann O., Pirner H.J. Z.Phys. {\bf C21} (1984) 277.
\vspace{-3mm}
\bibitem{efr10} Levin E.M., Ryskin M.G. Yad.Fiz. {\bf 41} (1985) 1622;
Brodsky S., Hoyer P. Nucl.Phys. {\bf A539} (1991) 79; {\bf B369} (1992) 519.
\vspace{-3mm}
\bibitem{efr11} Mueller A.H., Qiu J. Nucl.Phys. {\bf B264} (1986) 537.
\vspace{-3mm}
\bibitem{efr11a} Indumathi D. Prepr. DO-TH-96/17, hep-ph/9609361.
\vspace{-3mm}
\bibitem{efr12} Llewellin Smith C.H. Phys.Lett. {\bf B128} (1983) 107;
Ericson M., Thomas A.W. Phys.Lett. {\bf B128} (1983) 112;
Titov A.I. Yad.Fiz. {\bf 40} (1984) 76;
Akulinichev et al. Phys.Lett. {\bf B158} (1985) 485;
Pis'ma JETP {\bf 42} (1985) 105; Phys.Rev.Lett. {\bf 55} (1985) 2239;
Birbrair et al. Phys.Lett. {\bf B166} (1986) 119;
Saperstein E.E., Shmatikov M. Pis'ma JETP {\bf 41} (1985) 44.
\vspace{-3mm}
\bibitem{efr12a} Smirnov G.I. Phys.Lett. {\bf B364} (1995) 87;
hep-ph/9512204.
\vspace{-3mm}
\bibitem{efr13} Morley P.D., Schmidt I. Phys.Rev. {\bf D34} (1986) 1305;
Berger E.L., Coester F., Wiringa R.B. Phys.Rev. {\bf D29} (1984) 398.
\vspace{-3mm}
\bibitem{efr14} Frankfurt L.L., Strikman M.I. Nucl.Phys. {\bf B148} (1982)
107.
\vspace{-3mm}
\bibitem{efr15} Szwed J. Phys.Lett. {\bf B128} (1983) 245.
\vspace{-3mm}
\bibitem{efr16} Jaffe R.L. Phys.Rev.Lett. {\bf 50} (1983) 228;
Date S. Progr.Theor.Phys. {\bf 70} (1983) 1682;
Carlson C.E., Havens T.Y. Phys.Lett. {\bf 51} (1983) 261;
Titov A.I. Yad.Fiz. {\bf 40} (1984) 76;
Zotov N.P., Saleev V.A., Tsarev V.A. Pis'ma JETP {\bf 40} (1984) 200;
Jad.Fiz. {\bf 45} (1987) 561;
Chemtob M., Peschansky R. J.Phys.G. {\bf 10} (1984)  599;
Dias de Deus J., Varela M. Phys.Rev. {\bf D30} (1984) 697;
Bondarchenko E.A., Efremov A.V. Prepr. JINR E2-84-124 (1984);
Kondratyuk L.A., Shmatikov M. Z.Phys. {\bf A321} (1985) 301;
Yad.Phys. {\bf 41} (1985) 222;
Clark B.C. et al. Phys.Rev. {\bf D31} (1985) 617;
Nguyen Q.B. et al. Acta Phys.Austr. {\bf 57} (1985) 277.
\vspace{-3mm}
\bibitem{efr17} Hoodbhoy P., Jaffe R.L. Phys.Rev. {\bf D35} (1987) 113.
\vspace{-3mm}
\bibitem{efr16a} Braun M., Vecherin V. Nucl.Phys. {\bf B427} (1994) 614;
see also hep-ph/9612237.
\vspace{-3mm}
\bibitem{efr18} Frankfurt L.L., Strikman M.I. EPAN {\bf 11} (1980) 571;
Phys.Rep. {\bf 76} (1981) 215.
\vspace{-3mm}
\bibitem{efr19} Arneodo et al. (EMC) Z.Phys. {\bf C35} (1987) 433.
\vspace{-3mm}
\bibitem{efr20} Savin I.A. Proc. 22-nd Intern.Conf. on HEP,
Leipzig, 1984, p.251.; Savin I.A., Smirnov G.I. EPAN {\bf 22} (1991) 1005.
\vspace{-3mm}
\bibitem{efr20a} Gorenshtein M.I., Zinoviev G.M., Shelest V.P. Yad. Fiz.
{\bf 26} (1977) 788;
Kalinkin B.N., Shmonin V.L. EPAN {\bf 11} (1980) 630;
Shuryak E.V. Yad. Fiz. {\bf 24} (1976) 620;
Kopeliovich V.B.,  Phys.Rep. {\bf 199} (1986) 51.
\vspace{-3mm}
\bibitem{efr21} Blokhintsev D.I. JETP {\bf 33} (1957) 1295.
\vspace{-3mm}
\bibitem{efr23} Azhgirey L.S. et al. JINR Prepr. P1-86-728 (1986);
Yad.Fiz. {\bf 46} (1987) No.9.
\vspace{-3mm}
\bibitem{efr22} Efremov A.V., Kaidalov A.B., Kim V.T., Lykasov G.I.,
Slavin N.V. Sov. Journ. Nucl. Phys. {\bf 47} (1988) 868.
\vspace{-3mm}
\bibitem{efr24} Leksin G.A. Proc. of 8-th Int. Sem. on Problems of HEP,
JINR D1,2-86-668, 1986, 259.
\vspace{-3mm}
\bibitem{efr25} Baldin A.M. et al. JINR Communication E1-82-472, 1982.
\vspace{-3mm}
\bibitem{efr25c} Bojarinov S.B. et al. Yad. Fiz. {\bf 50} (1989) 1605.
\vspace{-3mm}
\bibitem{efr25a} Belyaev I.M. e.a. Phys.Atom.Nucl. {\bf 56} (1993) 1378;
Gavreshuk O.P., Peresedov V.F., Zolin L.S.
Nucl.Phys. {\bf A523} (1991) 589.
\vspace{-3mm}
\bibitem{efr26} Arnold et al. Phys.Rev.Lett. {\bf 52} (1984) 727;
SLAC-PUB-3257 (1984).
\vspace{-3mm}
\bibitem{efr27} Bojarinov S.B. et al. Yad. Fiz. {\bf 54} (1991) 119
(Phys.Atom.Nucl. {\bf 54} (1991) 71); Yad. Fiz. {\bf 56} (1993) 125
(Phys.Atom.Nucl. {\bf 56} (1993) 72).
\vspace{-3mm}
\bibitem{efr28} Efremov A.V., Kaidalov A.B., Lykasov G.I.,
Slavin N.V. Yad.Fiz. {\bf 57} (1994) 932 (Phys.Atom.Nucl.
{\bf 57} (1993) 874).
\end{thebibliography}
\end{document}